\documentclass[twocolumn,amssymb, nobibnotes, aps, superscriptaddress]{revtex4-1}
\usepackage{graphicx} 
\usepackage{booktabs}
\usepackage{natbib}
\usepackage{color}
\usepackage[colorlinks=true, allcolors=blue]{hyperref}
\usepackage{hyperref}
\usepackage{amsmath}
\usepackage{bm}
\usepackage{comment}

\begin{document}
\title{Effect of mechanical strain on the optical properties of nodal-line semimetal ZrSiS}%
\author{Weiqing Zhou}%
\affiliation{Key Laboratory of Artificial Micro- and Nano-structures of Ministry of Education and School of Physics and Technology, Wuhan University, Wuhan 430072, China}
\author{A. N. Rudenko}
\email{a.rudenko@science.ru.nl}
\affiliation{Key Laboratory of Artificial Micro- and Nano-structures of Ministry of Education and School of Physics and Technology, Wuhan University, Wuhan 430072, China}
\affiliation{\mbox{Institute for Molecules and Materials, Radboud University, Heijendaalseweg 135, NL-6525 AJ Nijmegen, The Netherlands}}
\affiliation{\mbox{Theoretical Physics and Applied Mathematics Department,
Ural Federal University, 620002 Ekaterinburg, Russia}}
\author{Shengjun Yuan}%
\email{s.yuan@whu.edu.cn}
\affiliation{Key Laboratory of Artificial Micro- and Nano-structures of Ministry of Education and School of Physics and Technology, Wuhan University, Wuhan 430072, China}
\date{\today}
\begin{abstract}
Optical properties of nodal-line semimetal ZrSiS are studied using first-principles calculations. 
Frequency-independent optical conductivity is a fingerprint of the infrared optical response in ZrSiS. We find that this characteristic feature is robust with respect to uniaxial compressive strain of up to 10 GPa, yet with the flat region being narrowed with increasing strain. Upon uniaxial tensile stress of 2 GPa, the Fermi surface undergoes a Lifshitz transition accompanied by a weakening of the interband screening, which reduces the spectral weight of infrared excitations. 
We also show that the high-energy region is characterized by low-loss plasma excitations at $\sim$20 eV with essentially anisotropic dispersion. Strongly anisotropic dielectric properties suggest the existence of a hyperbolic regime for plasmons in the deep ultraviolet range. Although the frequencies of high-energy plasmons are virtually unaffected by external uniaxial deformation, their dispersion can be effectively tuned by strain.

\end{abstract}

\maketitle

\section{Introduction}
As three-dimensional analogues of graphene, Dirac and Weyl semimetals have attracted considerable attention in the last years \cite{fisher2019topological,PhysRevLett.108.140405,liu2014discovery,lv2015experimental,xu2015discovery}. Both Dirac and Weyl materials are characterized by linearly dispersing valence and conduction bands that cross at discrete point in momentum space, giving rise to low-energy excitations behaving like Dirac or Weyl fermions. 
Recently, a novel class of topological materials, nodal-line materials, has been predicted \cite{burkov2011weyl,burkov2011topological}. In comparison to Dirac and Weyl semimetals, band crossing in nodal-line semimetals occurs along continuous lines. Since 2011, several materials were proposed to be nodal-line semimetals  \cite{yu2015topological,chen2015topological,okamoto2016low}, and some of them have been confirmed experimentally using such techniques as angle-resolved photoemission spectroscopy (ARPES) \cite{bian2016topological,schoop2016dirac,topp2016non}, magnetotransport \cite{emmanouilidou2017magnetotransport,matusiak2017thermoelectric,hu2016evidence}, and optical \cite{Shao1168} measurements.

The family of ternary compounds ZrSi$X$ ($X$=S, Se, Te) is a typical example of nodal-line semimetal with well separated Dirac cones \cite{PhysRevB.95.161101,schoop2016dirac}. The presence of topologically nontrivial linear bands in ZrSi$X$ has been observed experimentally by several methods, including ARPES \cite{schoop2016dirac,topp2017surface,chen2017dirac,fu2017observation}, scanning probe techniques \cite{Butler2017Quasi,lodge2017observation}, as well as thermoelectric \cite{matusiak2017thermoelectric} and magnetotransport \cite{pezzini2018unconventional,hu2017nearly,wang2016evidence,singha2017large,ali2016butterfly,pan2018three} measurements of quantum oscillations. Among ZrSi$X$s, ZrSiS is especially prospective material for optoelectronic applications due to its high carrier mobility \cite{matusiak2017thermoelectric,sankar2017crystal}, thermal stability \cite{lam1997new}, and non-toxic nature \cite{neupane2016observation}. A significant attention has been paid to ZrSiS due to its unusual properties observed in experiment. 
Particularly, ARPES experiments reveal that ZrSiS hosts two kinds of nodal lines. While in the first kind the degeneracy of Dirac points is protected by non-symmorphic symmetry, in the second kind the degeneracy is lifted by the spin-orbit coupling, inducing a small gap of the order of 10 meV \cite{schoop2016dirac}.
The upper limit of this gap ($\sim$ 30~meV) is observed by recent low-frequency optical measurements \cite{schilling2017flat}. Compared to other known 3D Dirac materials, the energy range of the linearly dispersing bands in ZrSiS reaches 2 eV, making this material a promising candidate for studying Dirac fermions. Apart from Dirac physics, extremely strong Zeeman splitting with a large $g$-factor has been observed by measuring de Haas-van Alphen (dHvA) oscillations \cite{hu2017nearly}. 
There is also evidence of an important role of the correlation effects in ZrSiS and related materials.
The unusual mass enhancement of charge carriers in ZrSiS has been recently observed experimentally at low-temperatures \cite{pezzini2018unconventional}, which can be understood in terms of unconventional electron-hole pairing \cite{rudenko2018excitonic,scherer2018}. Last but not least, recent high-pressure electrical transport measurements pointed to the possibility of a topological phase transition in ZrSiS below 0.5 GPa \cite{vangennep2019possible}.

In comparison to conventional metals, Dirac semimetals have raised intense interest both from fundamental and applied perspectives due to their intriguing optical properties \cite{fisher2019topological,kuzmenko2008universal,wang2019anomalous}.
Recently, optical spectra of ZrSiS were measured in a large frequency range, from the near-infrared to the visible \cite{schilling2017flat}. It was found that the absorption spectrum remain almost unchanged for photon energies in the range from 30~meV to 350~meV \cite{schilling2017flat}. 
As has been pointed out by B\'acsi and Virosztek \cite{PhysRevB.87.125425}, in a noninteracting electron system with two symmetric
energy bands touching each other at the Fermi level, the real part of the interband optical conductivity $\sigma_{1}(\omega)$ demonstrates a power-law frequency dependence with $\sigma_{1} \propto (\frac{\hbar\omega}{2})^{(d-2)/z}$, where $d$ and $z$ are the dimension of the system and the power law of the band dispersion, respectively \cite{PhysRevB.87.125425}.
The flat optical conductivity is typical for graphene ($d$=2 and $z$=1), being a universal constant for Dirac electrons in two dimensions \cite{kuzmenko2008universal,mak2008measurement}.
In three dimensions, this behavior is not universal. Linear dependence is reported in point-node Dirac or Weyl semimetals as ZrT$_{5}$ \cite{PhysRevB.92.075107}, TaAs \cite{PhysRevB.93.121110}, and Cd$_{3}$As$_{2}$ \cite{PhysRevB.93.121202} ($d$=3 and $z$=1).
The flatness of the optical conductivity in ZrSiS is determined by an appropriate combination of intraband and interband transitions \cite{habe2018dynamical}.
Followed by the flat region, the optical conductivity in ZrSiS exhibits a characteristic U-shape ending at a sharp peak around 1.3 eV \cite{schilling2017flat,ebad2019chemical}. Interestingly, the optical response is strongly anisotropic with the 1.3 eV peak appearing in the in-plane [100] direction only \cite{habe2018dynamical}. Besides, essentially anisotropic magnetoresistance in ZrSiS has been measured experimentally \cite{lv2016extremely,wang2016evidence}. Recent findings on the family of compounds ZrSi$X$ ($X$=S, Se, Te) and ZrGe$X$ ($X$=S, Te) suggest that their optical properties are closely connected to the interlayer bonding, and can be tuned by external pressure \cite{ebad2019chemical}.

Unlike infrared and visible spectral regions, ultraviolet optical response of ZrSiS has not been studied yet. 
Besides that, previous works focus on the optical properties of pristine ZrSiS, while effect of strain, has not been addressed in detail. 
The ultraviolet region is especially appealing for plasmonic applications, for which ZrSiS appears promising due to its high carrier mobility, closely related to the sustainability of plasmonic modes. 
Short propagation length (lifetime) of plasmons in typical plasmonic materials (e.g., noble metals) represents a bottleneck for applications \cite{politano2015influence}.
At the same time, the application domain of ultraviolet plasmonics is highly diverse. It includes biochemical sensing applications \cite{mcmahon2013plasmonics,taguchi2012tailoring}, photodetection \cite{dhanabalan2016present}, nano-imaging \cite{zhang2015ultraviolet}, material characterization \cite{nakashima2004deep}, and absorption of radiation \cite{hedayati2014plasmonic}.

In this paper, we study broadband optical properties of ZrSiS crystals with a special emphasis on the effect of external strain. To this end, we use first-principles calculations in combination with the random phase approximation for the dielectric screening. We find that although the low-energy optical conductivity remains frequency-independent under uniaxial loading of up to 10 GPa, the corresponding spectral region is narrowing with increasing stress.
In the presence of tension, we observe an electronic Lifshitz transition at around 2 GPa. This transition results in a suppressed intraband screening, which reduces the spectral weight in the infrared region.
Apart from the flat optical conductivity at low energies, our calculations show that ZrSiS is characterized by high-energy plasma excitations with frequencies around 20 eV.
Given that the optical response in ZrSiS is highly anisotropic, it permits the existence of low-loss hyperbolic plasmons in the ultraviolet spectral range.  

The paper is organized as follows. In Sec.~II, we describe our computational method and calculation details. Optical properties of pristine ZrSiS are presented in Sec.~III, where we specifically focus on the low- and high-energy spectral regions. In Sec.~IV, we study the effect of external strain on the optical conductivity and plasma excitations in ZrSiS. In Sec.~V, we summarize our findings.

\section{Calculation details}

\subsection{Electronic structure}
\begin{figure*}[tbp]	
	\includegraphics[width=18cm]{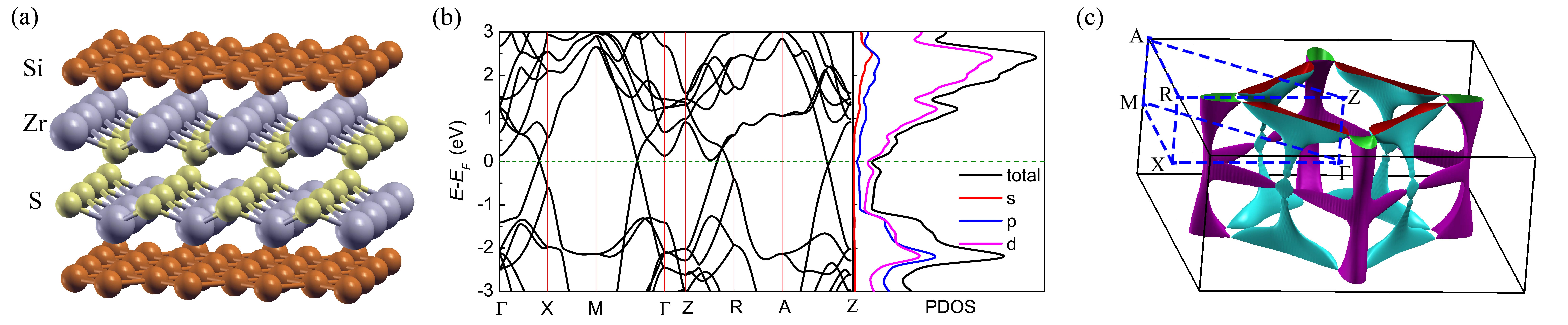}
	\caption{(a) Schematic representation of the ZrSiS crystal structure; (b) Calculated band structure and orbital-resolved density of states in the vicinity of the Fermi energy; (c) Three-dimensional view of the Fermi surface with purple and cyan colors denoting valence and conduction states, respectively. Black lines mark the Brillouin zone boundaries. Dashes blue lines connect the high-symmetry points used in (b). }
	\label{fig:label1}
\end{figure*}

ZrSiS is a layered crystal with a tetragonal structure and space group P4/$nmm$ (No.129). Its structure is formed by Zr-S layers sandwiched between Si layers, and periodically repeated in the direction normal to the layers, as shown in Figure~\ref{fig:label1}(a). The equilibrium lattice constants obtained from full structural optimization at the DFT level are $a=3.56$~\AA~(in-plane) and $c=8.17$ \AA ~(out-of-plane). 
The DFT electronic structure calculations are performed within the pseudopotential plane-wave method as implemented in {\sc quantum espresso} \cite{giannozzi2017advanced} simulation package. We use generalized gradient approximation (GGA) \cite{perdew1996generalized} in combination with norm-conserving pseudopotentials \cite{hamann1979norm}, in which $4s$ and $4d$ electrons of Zr, $3s$ and $3p$ electrons of Si, as well as $3s$ and $3p$ electrons of S were treated as valent. The reciprocal space was sampled by a uniform ($24\times24\times8$) {\bf k}-point mesh. In the calculations, we set the energy cutoff for the plane-wave basis to 80 Ry, and a self-consistency threshold for the total energy to $10^{-12}$ Ry. The atomic structure and lattice parameters were optimized until the residual forces on each atom were less than $10^{-5}$ Ry/Bohr. The effect of spin-orbit coupling is not taken into account in our study as it is only relevant for low temperatures ($<100$ K) and in the low-frequency region ($< 20$~meV) \cite{schilling2017flat}. 
All crystal graphics was generated by means of {\sc xcrysden} visualization package \cite{kokalj2003computer}.

\subsection{Dielectric function}
Dielectric function $\epsilon(\mathbf{q},\omega)$ was calculated within the random phase approximation (RPA) using {\sc yambo} \cite{marini2009yambo} package. 
Its standard form as function of wave vector $\mathbf{q}$ and frequency of incident photon $\omega$ reads:
\begin{equation}
	\epsilon(\mathbf{q},\omega)=1-v(\mathbf{q})\chi^{0}(\mathbf{q},\omega),
\label{dielec}
\end{equation}
where $v(\mathbf{q})=\frac{4\pi e^2}{|\mathbf{q}|^{2}}$ is the bare Coulomb potential, 
$\chi^{0}$ is the irreducible response function evaluated within the independent particle approximation \cite{marini2009yambo}:
\begin{multline}
\chi^{0}(\mathbf{q},\omega) = \frac{2}{V} \sum_{{\bf k},nm} \rho^{*}_{nm\mathbf{k}}(\mathbf{q})\rho_{nm\mathbf{k}}(\mathbf{q})
 \\
\times \left[\frac{f_{n\mathbf{k}-\mathbf{q}}(1-f_{m\mathbf{k}})}{\omega+\varepsilon_{n\mathbf{k}-\mathbf{q}}-\varepsilon_{m\mathbf{k}}+i\eta}-\frac{f_{n\mathbf{k}-\mathbf{q}}(1-f_{m\mathbf{k}})}{\omega+\varepsilon_{m\mathbf{k}}-\varepsilon_{n\mathbf{k}-\mathbf{q}}-i\eta}\right],
\label{inter}
\end{multline}
where 
\begin{equation}
\rho_{nm\mathbf{k}}(\mathbf{q})=\langle n\mathbf{k}|e^{i\mathbf{q}\cdot\mathbf{r} } |m\mathbf{k}-\mathbf{q} \rangle
\label{rho}
\end{equation}
is the dipole transition matrix element, 
$f_{n\mathbf{k}}$ is the Fermi occupation factor, for which $T=300$ K was used in all calculations, $|n\mathbf{k}\rangle$ is the Bloch eigenstate corresponding to the band $n$ and wave vector $\mathbf{k}$, and $V$ is the cell volume.
To avoid computationally demanding calculations, we assume the scalar form of $\epsilon({\bf q}, \omega)$ and $\chi^0({\bf q},\omega)$, meaning that only ${\bf G}=0$ and ${\bf G}'=0$ elements of the full matrices are calculated. Physically, this approximation corresponds to the situation, in which the local field effects are neglected, i.e. $\epsilon({\bf r}_1,{\bf r}_2) \simeq \epsilon(|{\bf r}_1-{\bf r}_2|)$. This approximation is well justified for 3D systems with weak inhomogeneities of the charge density \cite{onida2002electronic}. 
In Eq.~(\ref{inter}), $\eta$ is the damping parameter playing the role of the electron linewidth, which can be attributed to the imaginary part of the self-energy, $\eta 
\sim \mathrm{Im}[\Sigma(\omega,\mathbf{k})]$ \cite{marder2010condensed}. Here, we do not detail the scattering mechanism and consider $\eta$ as a free parameter.

To reproduce the quantities measured in optical experiments, one needs to evaluate the long-wavelength limit of the dielectric function,
\begin{equation}
\epsilon_{\alpha \alpha}(\omega)\equiv \lim_{\mathbf{q} \to 0} \epsilon(\mathbf{q},\omega),
\end{equation}
where $\alpha$ is the direction of the incident light, and the limit is taken with ${\bf q}$ parallel to $\alpha$.
Taking this limit numerically is a computationally nontrivial task as it requires high density of {\bf q}-point to be included in the calculations. This can be avoided by expanding the dipole transition matrix elements at $\mathbf{q} \to 0$ using $e^{i\mathbf{q}\cdot\mathbf{r} } \approx 1 +i \mathbf{q}\cdot\mathbf{r}$.
To this end, the matrix elements $\mathbf{r}_{nm\mathbf{k}}=\langle n \mathbf{k}|\mathbf{r}|m\mathbf{k} \rangle$ needs to be computed. Within the periodic boundary conditions using the relation $[\mathbf{r},H]=\mathbf{p}+[\mathbf{r},V_{nl}]$ one arrives at \cite{sangalli2019many}
\begin{equation}
\langle n \mathbf{k}|\mathbf{r}|m\mathbf{k} \rangle = \frac{\langle n \mathbf{k}|\mathbf{p}+[\mathbf{r},V_{nl}]|m\mathbf{k} \rangle}{\varepsilon_{n\mathbf{k}}-\varepsilon_{m\mathbf{k}}}
\end{equation}
where $V_{nl}$ is the nonlocal part of the pseudopotential. 

At ${\bf q}\to 0$, Eq.~(\ref{inter}) does not explicitly takes intraband transitions into account. Since ZrSiS is a semimetal, the intraband transition provide an important contribution to the dielectric response at low energies. To account for this contribution, we calculate the Drude corrections to the dielectric function $\epsilon^{\mathrm{intra}}_{\alpha \alpha}(\omega)=\epsilon^{\mathrm{intra}}_{1,\alpha \alpha}(\omega)+i\epsilon^{\mathrm{intra}}_{2,\alpha \alpha}(\omega)$, which are evaluated from the standard free-electron plasma model \cite{dressel2002electrodynamics}:
\begin{equation}
\begin{aligned}
\epsilon^{\mathrm{intra}}_{1,\alpha \alpha}(\omega)=1-\frac{\omega_{p,\alpha \alpha}^{2}}{\omega^{2}+\delta^{2}}, \\
\epsilon^{\mathrm{intra}}_{2,\alpha \alpha}(\omega)=\frac{\delta\omega_{p,\alpha \alpha}^{2}}{\omega^{3}+\omega\delta^{2}}.
\label{Drude_eps}
\end{aligned}
\end{equation}
Here, $\delta$ has similar physical meaning as $\eta$ in Eq.~(\ref{inter}), and $\omega_{p,\alpha \alpha}$ is the $\alpha$-component of the (unscreened) plasma frequency given by \cite{lee1994first,harl2007ab}:
\begin{equation}
	\omega^{2}_{p,\alpha\beta}=-\frac{4\pi e^{2}}{V}\sum_{n,\mathbf{k}}\frac{\partial f_{n\mathbf{k}}}{\partial \varepsilon_{n{\bf k}}} v^{\alpha}_{n{\bf k}} v^{\beta}_{n{\bf k}}
	\label{plasma}
\end{equation}
where 
$v^{\alpha}_{n{\bf k}}=\hbar^{-1}\partial \varepsilon_{n{\bf k}}/\partial k_{\alpha}$ 
is the $\alpha$-component of the group velocity of the electrons with wave vector ${\bf k}$ at band $n$. In this work, the plasma frequency is calculated using the {\sc simple} code \cite{prandini2019simple}.

The intraband contribution to the optical conductivity can be calculated accordingly, using the well-known expressions \cite{marder2010condensed}:
\begin{equation}
\begin{aligned}
\sigma^{\mathrm{intra}}_{1,\alpha\alpha}(\omega)=\frac{\omega \epsilon^{\mathrm{intra}}_{2,\alpha\alpha}(\omega)}{4\pi} \\
\sigma^{\mathrm{intra}}_{2,\alpha\alpha}(\omega)=1 - \frac{\omega \epsilon^{\mathrm{intra}}_{1,\alpha\alpha}(\omega)}{4\pi}
\end{aligned}
\end{equation}

\section{Optical properties of pristine \texorpdfstring{Z\MakeLowercase{r}S\MakeLowercase{i}S}{ZrSiS}}
\subsection{Low-energy region}
We first calculate the electronic structure of ZrSiS for its equilibrium crystal structure.  
In Figure~\ref{fig:label1}, we show the band structure, density of states projected on $s$-, $p$-, and $d$-orbitals (PDOS), and the corresponding Fermi surface. 
The most prominent feature of the band structure is a series of linearly dispersing bands with the Dirac-like crossings in the vicinity of the Fermi energy ($\varepsilon_F$). The linear bands extend over a rather large energy range of up to 2 eV. From Figure~\ref{fig:label1}(b), one can see that DOS exhibits a minimum at $\varepsilon_F$, as expected near the band crossing points. In the range from $-1$ to 0 eV, the valence states are entirely formed by linearly dispersed bands, while the states above $\varepsilon_F$ are mixed with quadratic bands, giving rise to a larger DOS for the conduction band.
As can be seen from PDOS, $d$-orbitals have dominant contribution to the states near $\varepsilon_F$. At $\varepsilon \lesssim1$ eV there is a comparable contribution from $p$-orbitals.
In Figure~\ref{fig:label1}(c), we show the corresponding Fermi surface. It is composed of two distinct parts, corresponding to electron (cyan) and hole (purple) states. Each part is formed by four disconnected pockets. As we will see below, the Fermi surface topology plays an important role in the optical properties of strained ZrSiS.

\begin{figure}[btp]	
	\includegraphics[width=8.5cm]{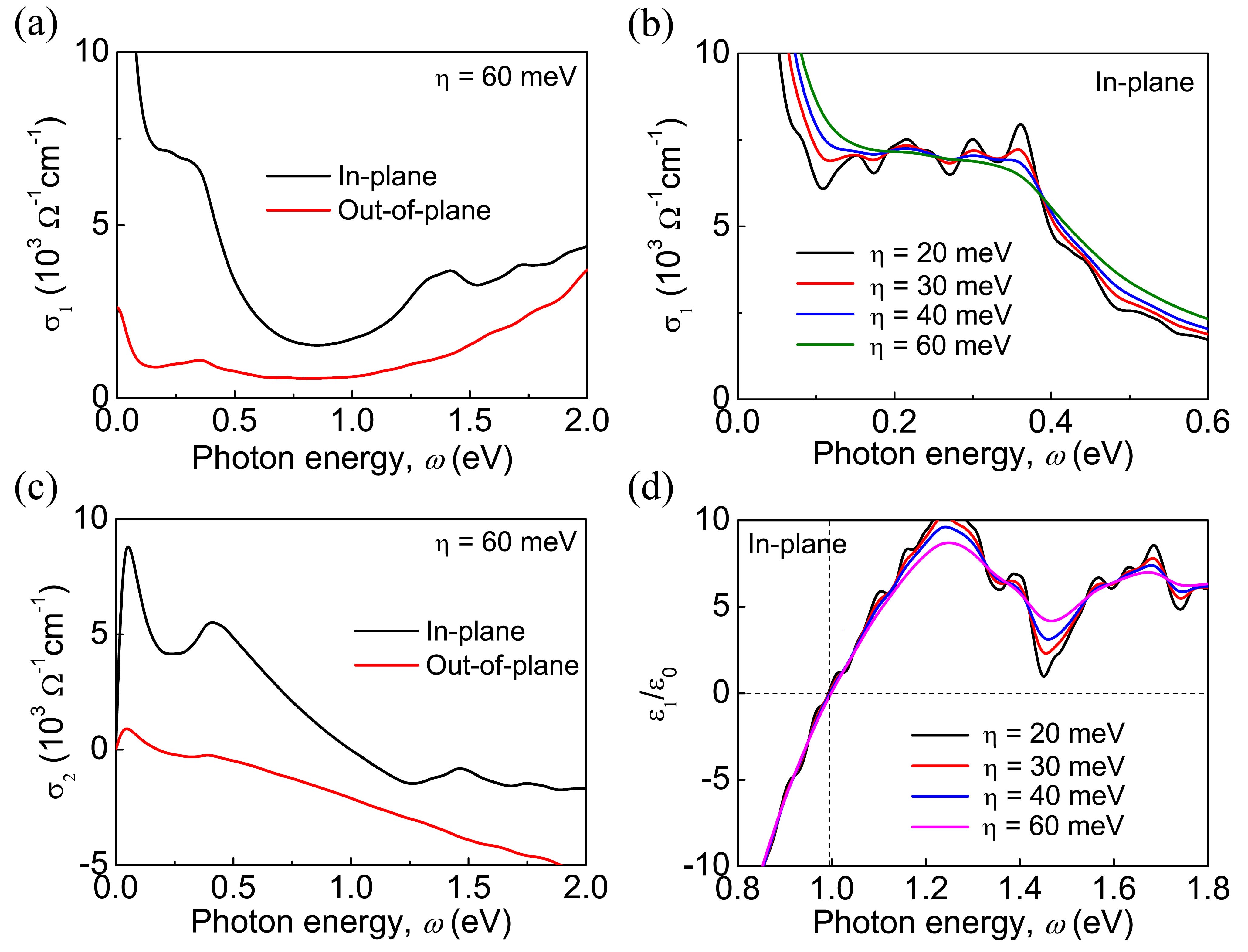}
	\caption{(a) Real part of the optical conductivity shown as a function of the photon energy with incidence along in-plane [100] and out-of-plane [001] crystallographic directions; (b) Real part of the in-plane optical conductivity calculated for different damping parameters $\eta$; (c) Imaginary part of the optical conductivity calculated along [100] and [001] directions; (d) Real part of the in-plane dielectric function calculated for different $\eta$.}
	\label{fig:label2}
\end{figure}

\begin{figure*}[tbp]	
	\includegraphics[width=13cm]{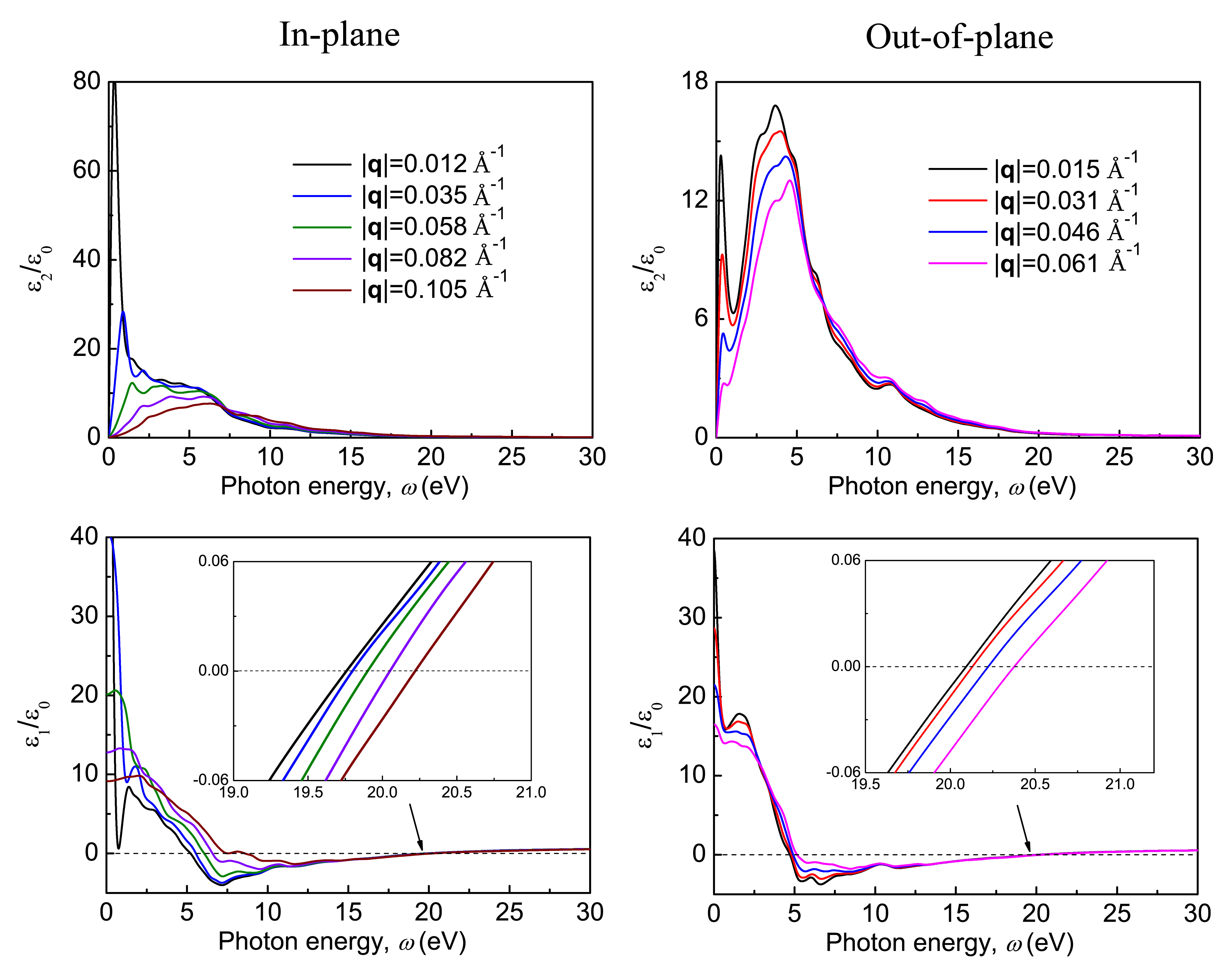}
	\caption{Imaginary (upper panels) and real (lower panels) parts of the dielectric function of pristine ZrSiS calculated as a function of the photon energy $\omega$ for a series of wave vectors $\mathbf{q}$ along the in-plane (left panels) and out-of-plane (right panels) directions. Inset shows a zoom-in of the high-energy region where $\epsilon_1({\bf q},\omega)=0$. }
	\label{fig:label3}
\end{figure*}

After the ground state electronic structure is obtained, we calculate the dielectric functions, and the corresponding optical conductivities. We start from the ${\bf q}\to 0$ limit and first calculate the unscreened plasma frequencies using Eq.~(\ref{plasma}). We arrive at $\omega_{p,xx}=3.15$ eV and $\omega_{p,zz}=1.08$ eV for the in-plane [100] and out-of plane [001] components, respectively. The value obtained for the [100] directions is in good agreement with the experimental estimate of 2.88 eV \cite{schilling2017flat}. 
In Figures~\ref{fig:label2}(a) and \ref{fig:label2}(c), we show the real and imaginary parts of the optical conductivity calculated in the region up to 2 eV for [100] and [001] directions of photon propagation. The spectral weight obtained for the in-plane direction is significantly larger compared to the out-of-plane direction. This indicates a strong anisotropy between the optical response in ZrSiS.

In order to assess sensitivity of the optical conductivity to the effects induced by finite electron linewidth, in Figure~\ref{fig:label2}(b) we show the real part of the low-energy optical conductivity calculated for different parameters $\eta$ at the range from 20 to 60 meV. 
From Figure~\ref{fig:label2}(b), one can clearly see the prominent flat conductivity from 0.1 to 0.4 eV. The flat conductivity $\sigma_{\mathrm{flat}}$ is estimated to be $\sim$7000 $\Omega^{-1}$cm$^{-1}$, which is good agreement with the experimental result of 6600 $\Omega^{-1}$cm$^{-1}$ \cite{schilling2017flat}. The flatness is well reproduced for $\eta=30$--$40$ meV, while larger values result in a noticeable smearing of the flat region. For $\eta \lesssim 20$ meV, one can see the emergence of an oscillatory behavior. This behavior is of the numerical origin, and can be associated with insufficient sampling of the Brillouin zone. In what follows, we set $\eta=40$~meV in all low-energy ($0$--$2$ eV) conductivity calculations.
This value is in agreement with the electron linewidth experimentally estimated in ZrSiS as $\sim$30 meV at 300 K \cite{schilling2017flat}. Following the flat region, there appears a U-shaped optical conductivity around $1.3$~eV \cite{ebad2019chemical}. The peak above the U-shaped region at $\sim$1.3~eV is only found for the in-plane direction, while it is absent in the out-of-plane direction. This peak mainly originates from the excitation between the linearly dispersing bands near $\varepsilon_{F}$ and from the transitions between quadratic bands in the direction from Z to R.

In Figure~\ref{fig:label2}(d), we show the real part of the calculated in-plane dielectric function. The condition $\epsilon_1(\omega^{\mathrm{scr}}_p)=0$ allows us to estimate the screened plasma frequency, which is found to be $\omega^{\mathrm{scr}}_{p}$ $\sim 1$ eV. Having determined $\omega^{\mathrm{scr}}_{p}$, we can estimate the effective screening induced by the interband transitions \cite{dressel2002electrodynamics}. The corresponding dielectric constant  $\epsilon_{\infty}=(\omega_{p}/\omega^{scr}_{p})^{2}\approx 9$, which is consistent with the experimental value of $\sim7.8$ \cite{schilling2017flat}.
To understand the effect of finite electron linewidth on $\omega^{scr}_{p}$, we also plot $\epsilon_{1}(\omega)$ for different parameters $\eta$ in Figure~\ref{fig:label2}(d).
Compared to the flatness of the optical conductivity, the screened plasma frequency is almost insensitive to $\eta$.

\subsection{High-energy region}
\begin{figure*}[ht]	
	\includegraphics[width=13cm]{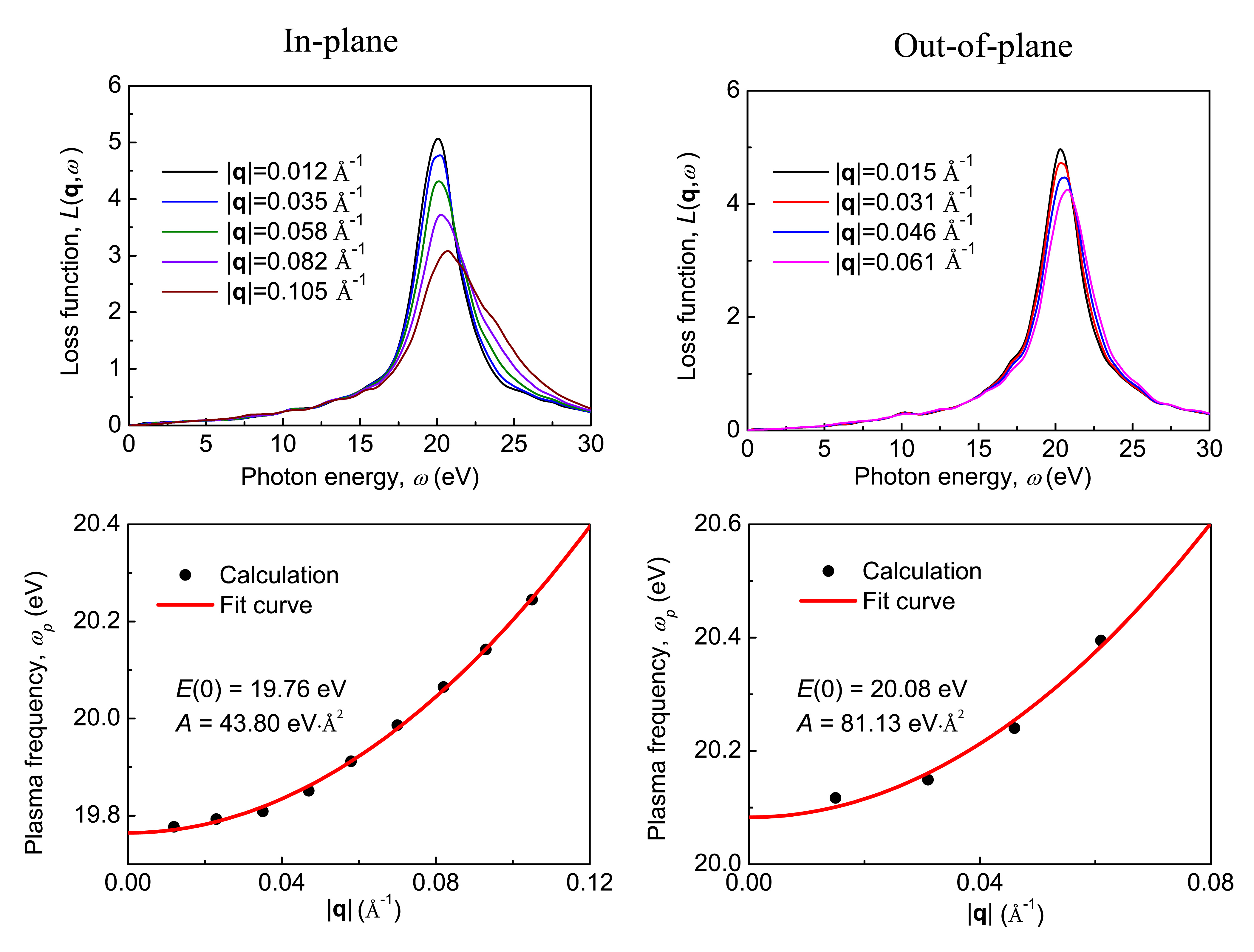}
	\caption{Upper panels: Electron energy loss spectrum $L({\bf q},\omega)$ as a function of the photon energy $\omega$ and momentum $\mathbf{q}$ calculated for the in-plane (left) and out-of-plane (right) directions. Lower panels: Dispersion of the high-energy plasmon $\omega_p({\bf q})$ calculated along the in-plane (left) and out-of-plane (right) directions.}
	\label{fig:label4}
\end{figure*}

We now turn to the optical response in the high energy region, $\omega>2$ eV. Here, we focus at the plasmonic excitations and consider momentum-resolved dielectric function $\epsilon({\bf q},\omega)$, which is shown in Figure~\ref{fig:label3} as a function of the photon energy for a series of small wave vectors ${\bf q}$ in both in-plane and out-of-plane directions. At $\omega \gtrsim$ 10 eV, $\epsilon({\bf q},\omega)$ is monotonic at small ${\bf q}$, with $\epsilon({\bf q},\omega) \rightarrow 1$ as $\omega \rightarrow \infty$, which is expected from the Drude model [Eq.~(\ref{Drude_eps})]. The most interesting energy region is determined by the condition $\epsilon_1(\mathbf{q},\omega)=0$, which defines the existence of plasma excitations. From Figure~\ref{fig:label3}, one can see that this criterion is fulfilled for two different energy regions: $\omega_p \sim 5$--7 and $\omega_p \sim 19$--20 eV.
To gain more insights in the plasmonic response, we calculate the energy loss function
\begin{equation}
L(\mathbf{q},\omega)=-\mathrm{Im}\left[\frac{1}{\epsilon(\mathbf{q},\omega)}\right],
\end{equation}
which can be associated with the Electron Energy Loss Spectroscopy (EELS) spectra.
Figure~\ref{fig:label4} shows $L({\bf q},\omega)$ calculated along the in-plane and out-of-plane directions of ZrSiS. In both cases, one can see a sharp peak around 20 eV, while there is no indication of the energy loss at lower energies. This means that the plasma oscillations around 5--7 eV are strongly damped. This can be understood from Figure~\ref{fig:label3}, where $\epsilon_2({\bf q},\omega)$ exhibits a peak around $\omega \sim 5$ eV, indicating strong absorption in this region. On the other hand, $\epsilon_2({\bf q},\omega)$ is almost zero around $\omega \sim $ 20 eV, indicating that high-energy plasmons are characterized by low losses, and could be observed experimentally. Recently, similar behavior has been experimentally observed in bulk black phosphorus crystal in the same frequency region \cite{nicotra2018anisotropic}.

The dispersion of bulk plasmons can be fitted with a second-order polynomial:
\begin{equation}
E(\mathbf{q}) = E(0) + A \, \mathbf{q}^{2},
\label{dispersion}
\end{equation}
where $E({0})$ is the plasmon energy at $\mathbf{q} \to {0}$ and $A$ is the dispersion coefficient. From Figure~\ref{fig:label4}, it can be seen that the calculated dispersion can indeed be fitted with Eq.~(\ref{dispersion}). Interestingly, although the plasma frequency is nearly independent of the direction of light propagation, the dispersion of high-energy plasmon modes is strongly anisotropic.
The existence of high-energy plasmons in ZrSiS might be beneficial in the context of ultraviolet optical devices \cite{sang2013comprehensive}. At the same time, strongly anisotropic dispersion of plasmon modes may give rise to unconventional plasma excitations, known as hyperbolic plasmons \cite{shekhar2014hyperbolic}.

\begin{figure}[ht]	
	\includegraphics[width=8.5cm]{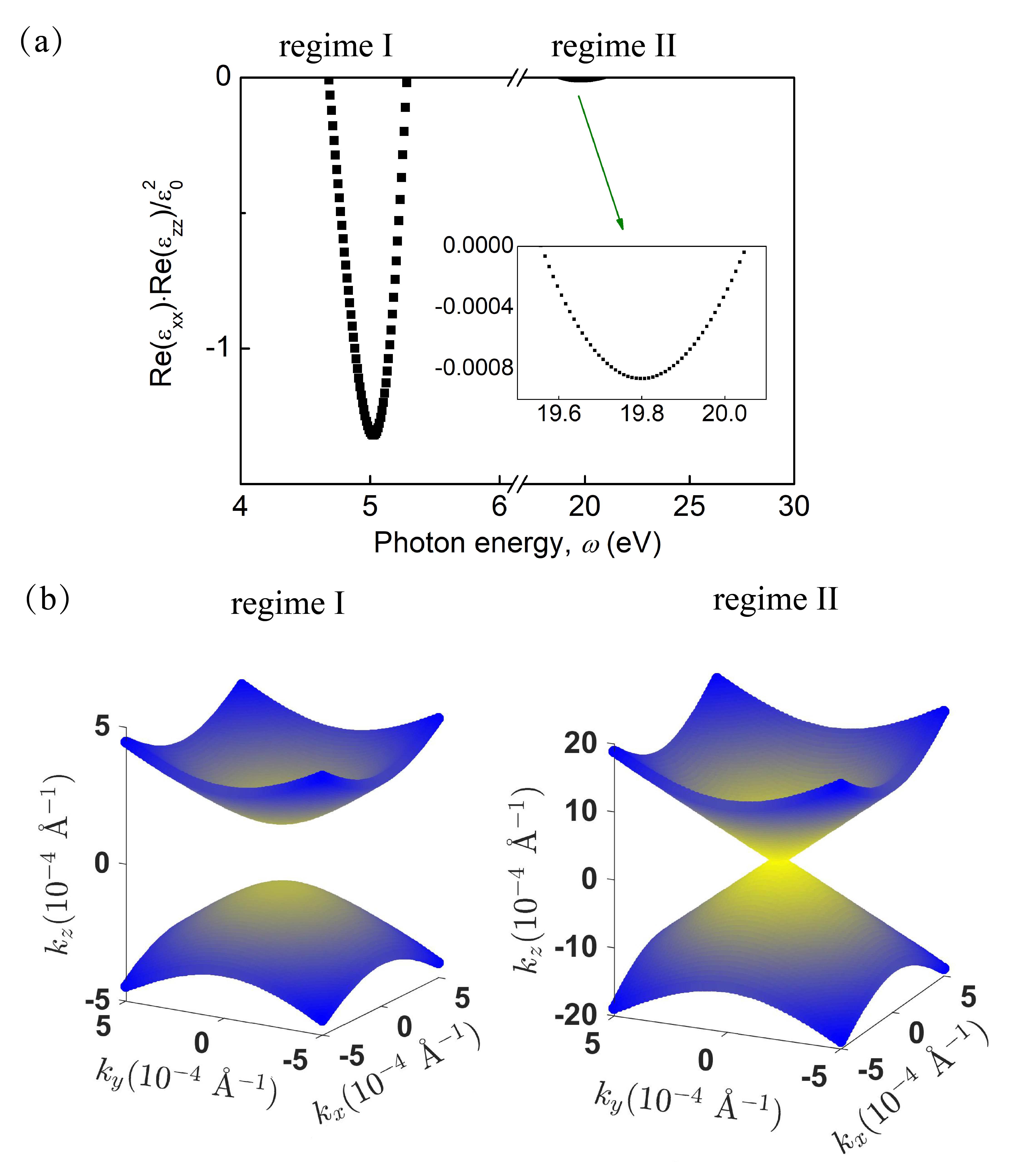}
	\caption{(a) Product of the in-plane and out-plane real dielectric functions shown as a function of energy; (b) Reciprocal-space representation of the constant-energy surfaces of two possible hyperbolic plasmon modes in ZrSiS, denoted as regime I (left) and II (right). The color shows the magnitude of $k_z$.}
	\label{fig:hyper}
\end{figure}

\begin{figure}[ht]	
	\includegraphics[width=9.0cm]{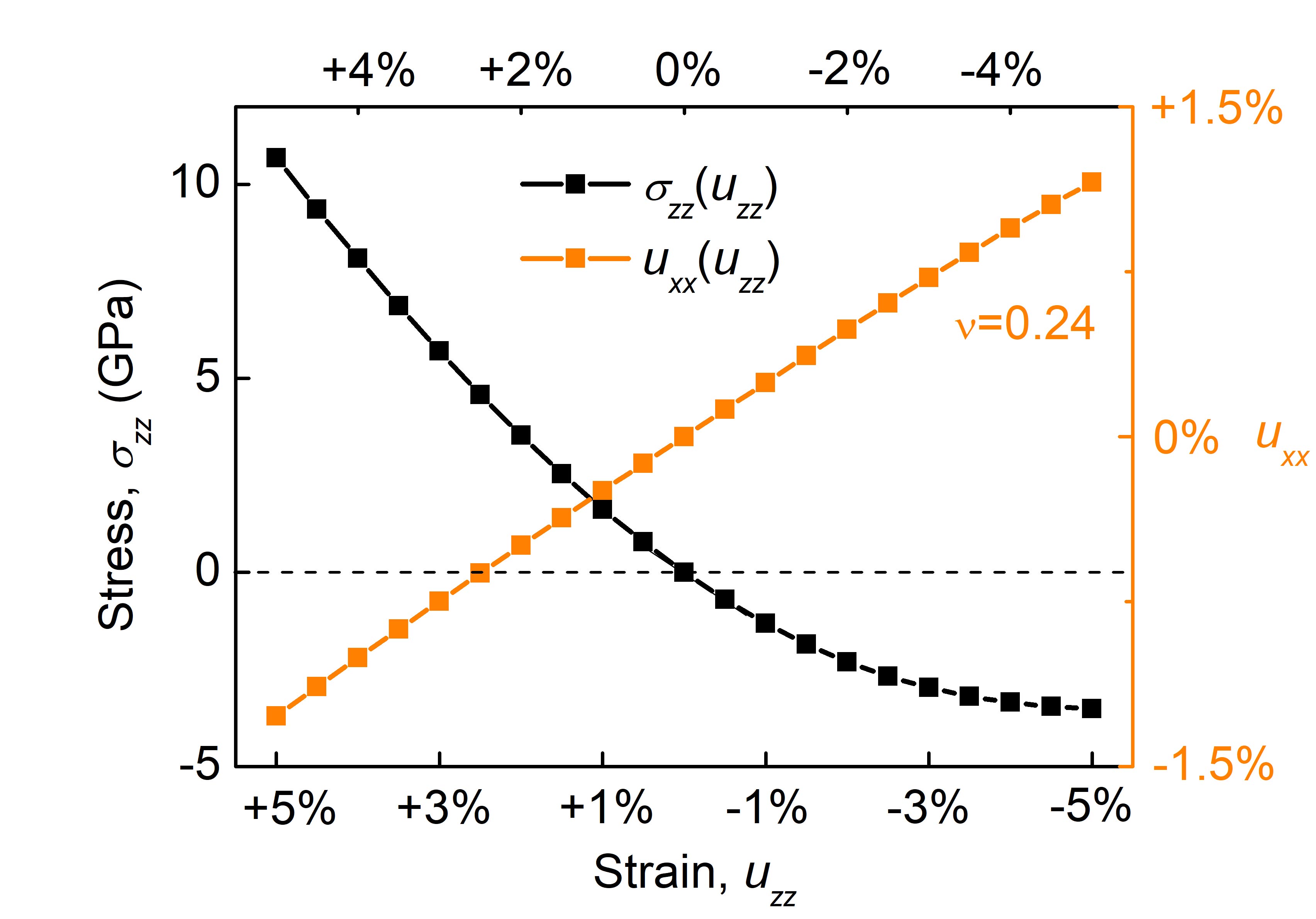}
	\caption{Black curve: The $zz$-component of the stress tensor ($\sigma_{zz}$) as a function of the uniaxial strain $u_{zz}$ in ZrSiS. Orange curve: In-plane strain $u_{xx}$ versus out-of-plane strain $u_{zz}$. $\nu=-\mathrm{d}u_{xx}/\mathrm{d}u_{zz}$ is the corresponding Poisson ratio estimated by linear regression.}
	\label{fig:stress}
\end{figure}

\begin{figure*}[ht]	
	\includegraphics[width=17cm]{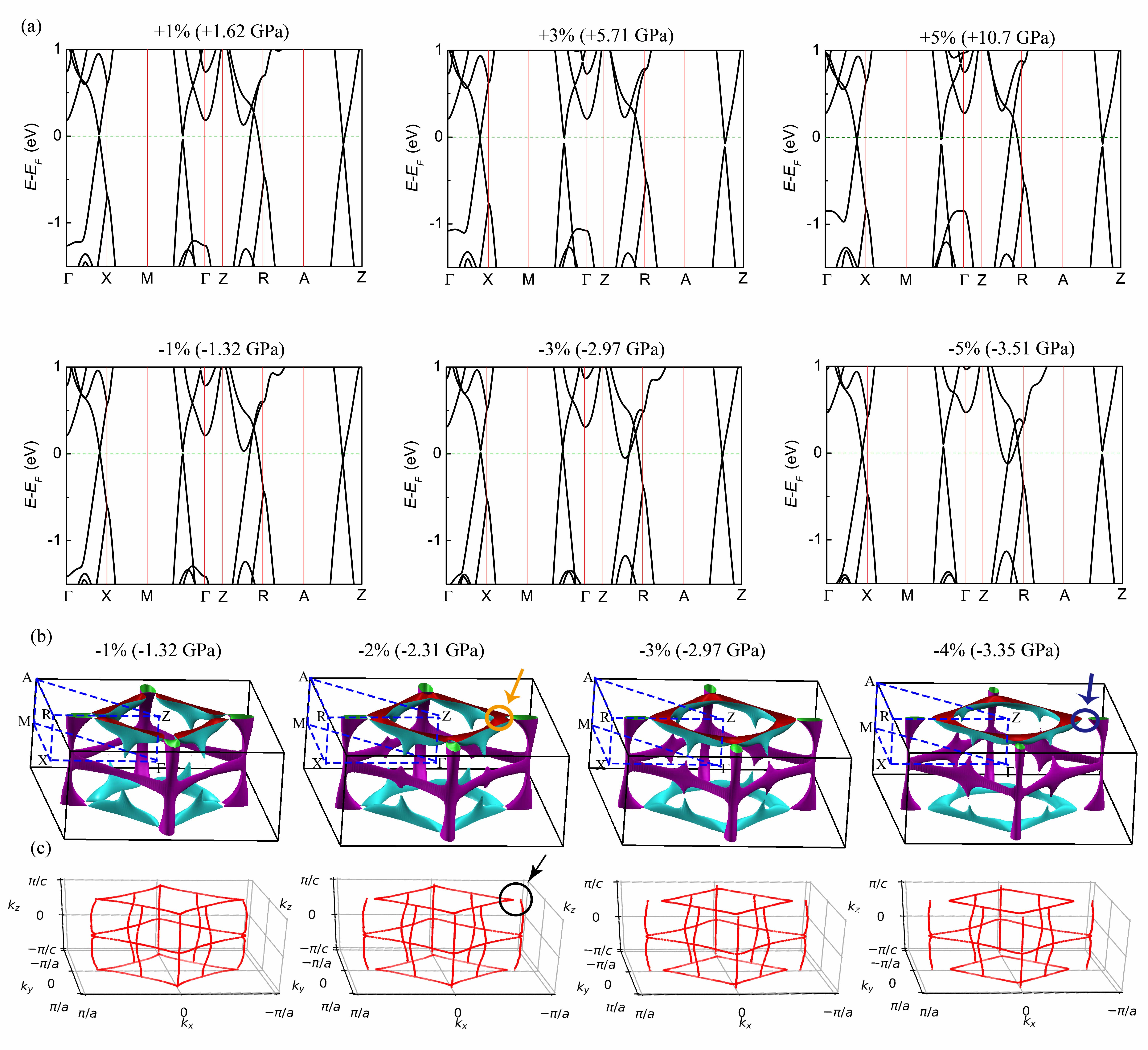}
	\caption{(a) Band structures calculated in the vicinity of the Fermi energy for different values of the uniaxial strain $u_{zz}$ in ZrSiS. Positive and negative values correspond to compression and tension, respectively. The related stress is given in parentheses;
	(b) The Fermi surfaces, and (c) corresponding nodal-line structure shown for the case of tensile strain, at which electronic Lifshitz transition is taking place. Circles and arrows highlight the location where the transition occurs.}
	\label{fig:label5}
\end{figure*}

Hyperbolic plasmons appear in crystals with strong anisotropy, in which effective permittivity changes sign with respect to the electric field direction \cite{gomez2015hyperbolic}.
The dispersion relation of light propagating in homogeneous layered material is determined by the relation:
\begin{equation}
	\frac{(k^{2}_{x}+k^{2}_{y})}{\epsilon_{zz}(\omega)}+\frac{k_{z}^{2}}{\epsilon_{xx}(\omega)}=\frac{\omega^{2}}{c^{2}},
\end{equation}
where $\epsilon_{xx}$ and $\epsilon_{zz}$ are the frequency-dependent permittivities along the in-plane and out-of-plane directions, respectively.
For frequencies at which  $\epsilon_{xx}(\omega)\cdot\epsilon_{zz}(\omega)<0$, the equation above describes a hyperboloid. This situation is considerably different from the closed spherical or elliptic dispersion typical for conventional materials with $\epsilon_{xx}(\omega)\cdot\epsilon_{zz}(\omega)>0$ \cite{gjerding2017layered,guo2012applications}. Depending on the form of the isofrequency surface, one can distinguish between the two types of hyperbolic materials: Type I if the hyperboloid is two-sheeted ($\epsilon_{zz}<0, \epsilon_{xx}>0$), and type II if the hyperboloid is single-sheeted ($\epsilon_{zz}>0, \epsilon_{xx}<0$).

In Figure~\ref{fig:hyper}(a), we show the corresponding permittivities calculated in ZrSiS as a function of the photon energy. One can see that the condition $\epsilon_{xx}(\omega)\cdot\epsilon_{zz}(\omega)<0$ is fulfilled in a narrow energy region around $\sim$5 eV and $\sim$20 eV, which are the frequencies at which the conventional bulk plasmon modes are found. In both cases, the hyperbolic plasmons may appear in a frequency range of about 0.6 eV.
Both hyperbolic modes demonstrate the dispersion relation of type I, corresponding to a two-sheeted hyperboloid, shown in Figure~\ref{fig:hyper}(b).
Simliar to other natural hyperbolic materials, hyperbolic regimes in ZrSiS appear only above the onset of intraband transition \cite{gjerding2017layered}. Since electronmagnetic waves propagating in hyprobolic materials follow the hyperbolic dispersion, hyprobolic media supports propagation of high-$\mathbf{k}$ waves that are evanescent in conventional media \cite{gomez2015hyperbolic}. Due to the properties of high-$\mathbf{k}$ waves, hyperbolic material have many potential applications, including negative refraction \cite{yao2008optical, hoffman2007negative}, sub-wavelength modes \cite{kapitanova2014photonic} and thermal emission engineering \cite{biehs2012hyperbolic}. We note, however, since the $\sim$5 eV mode is strongly damped, its practical significance is questionable.

\section{Optical properties of uniaxial strained \texorpdfstring{Z\MakeLowercase{r}S\MakeLowercase{i}S}{ZrSiS}}
Earlier studies on the family of compounds ZrSi$X$ ($X$=S, Se, Te) suggest that their physical properties are closely connected with the interlayer bonding. Moreover, the ratio of the out-of-plane and in-plane lattice constants $c/a$ can be considered as a measure for the interlayer bonding strength in these systems \cite{ebad2019chemical,ebad2019infrared,topp2016non}. In this regard, uniaxial strain applied in the out-of-plane direction is a promising way to tune the materials' properties.
Inspired by recent experimental works, which indicate the possibility of a topological phase transition in nodal-line semimetals under external pressure \cite{vangennep2019possible,ebad2019infrared}, here we study how the uniaxial strain would affect the optical properties of ZrSiS.

Before discussing the effect of strain on the electronic structure, we briefly focus on the mechanical properties of ZrSiS. We apply uniaxial strain in the direction perpendicular to the ZrSiS layers by varying the out-of-plane lattice constant $c$, and relaxing the in-plane lattice constant $a$. 
The stress is defined as $\sigma_{ij}=\frac{1}{\Omega}\frac{\partial F}{\partial u_{ij}}$, where $\Omega$ is the volume of unit cell, and $u_{ij}$ is the strain tensor. In our case, we focus on uniaxial strain assuming in-plane relaxation ($\sigma_{xx}=\sigma_{yy}=0$) and the absence of shear strain, i.e. $u_{xy}=u_{xz}=u_{yz}=0$. $F$ is the free energy of the crystal, which in the case of tetragonal symmetry (point group $D_{4h}$) is given by \cite{landau1989course}
\begin{multline}
    F=\frac{1}{2}\lambda_{xxxx}(u^{2}_{xx}+u^{2}_{yy})+\frac{1}{2}\lambda_{zzzz}u^{2}_{zz}+ \\
    \lambda_{xxzz}(u_{xx}u_{zz}+u_{yy}u_{zz})+\lambda_{xxyy}u_{xx}u_{yy},
\end{multline}
where $\lambda$ is the tensor of elastic moduli.
The calculated stress-strain curves are shown in Figure~\ref{fig:stress}. In case of uniaxial compressive strain along the out-of-plane direction ($u_{zz}$), the $\sigma_{zz}$ vs. $u_{zz}$ curve is nearly linear, indicating typical elastic regime and applicability of the Hooke's law. On the other hand, as can be seen from Figure~\ref{fig:stress}, the tensile  strain is highly nonlinear already at $2\%$ tension. The observed nonlinearity of the elastic properties indicates a considerable modification of the electronic structure upon tensile strain. In Figure~\ref{fig:stress}, we also show the dependence of the in-plane strain $u_{xx}$ with respect to $u_{zz}$. For $u_{zz}$ in the range from $-$5\% to $+$5\%, we obtain a perfect linear dependence, which allows us to estimate the Poisson's ratio. We obtain $\nu=-\mathrm{d}u_{xx}/\mathrm{d}u_{zz}=0.24$, which is in agreement with the results of previous studies \cite{2017Possion}.

\begin{figure}[tbp]	
	\includegraphics[width=8.5cm]{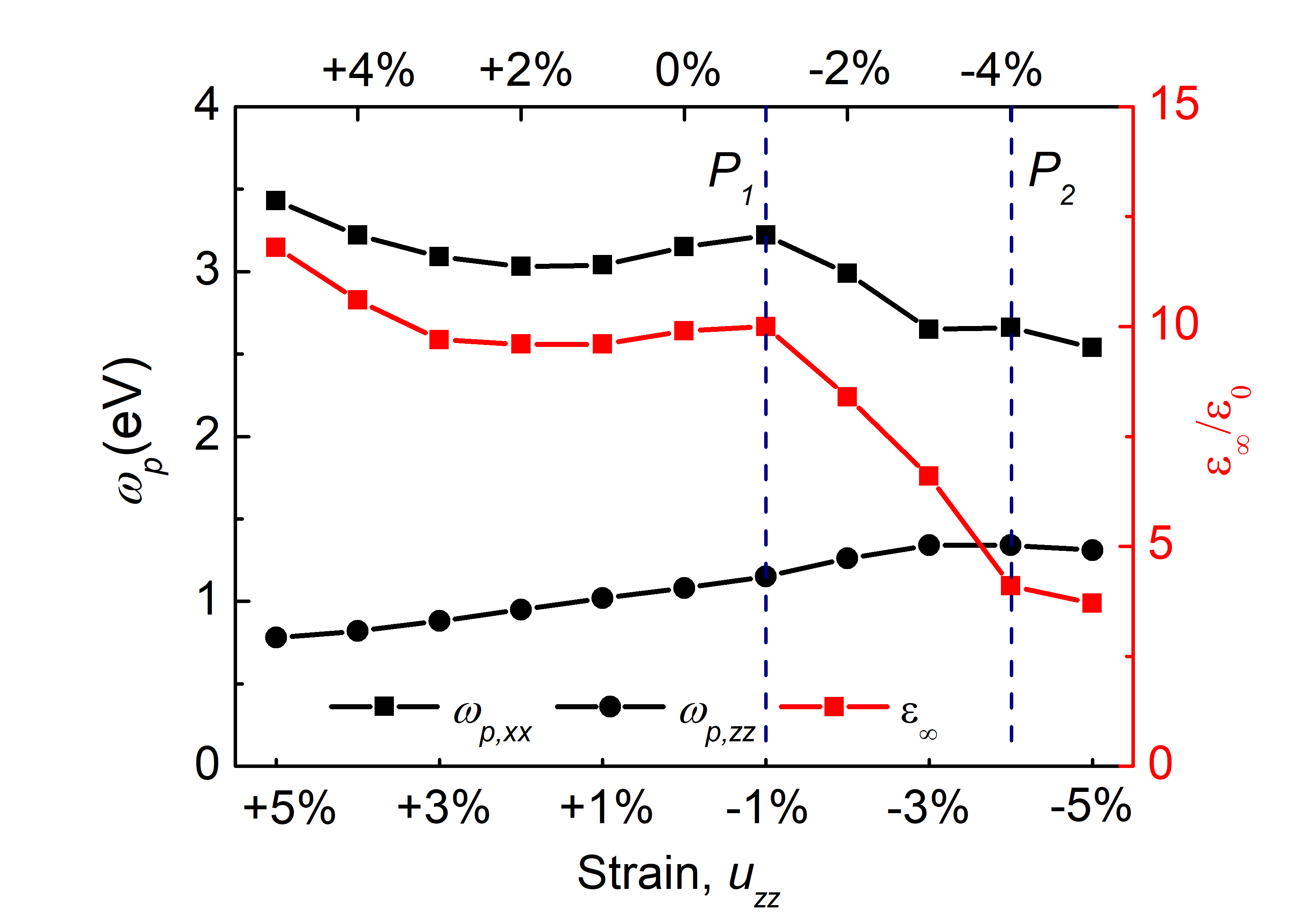}
	\caption{Strain-dependent low-energy plasma frequency $\omega_{p}$ (black) estimated using Eq.~(\ref{plasma}) for in-plane and out-of-plane directions, and intraband screening constant $\epsilon_{\infty}=(\omega_{p,xx}/\omega_{p,xx}^{scr})^2$ (red) calculated for the in-plane direction shown as a function of the uniaxial strain. Positive and negative values correspond to compression and tension, respectively.
	$P_{1}=-1.3$ GPa and $P_{2}=-3.4$ GPa are the critical stress values around which the electronic Lifshitz transition takes place. }
	\label{fig:strain_plasma}
\end{figure}

\begin{figure}[tbp]	
	\includegraphics[width=8.7cm]{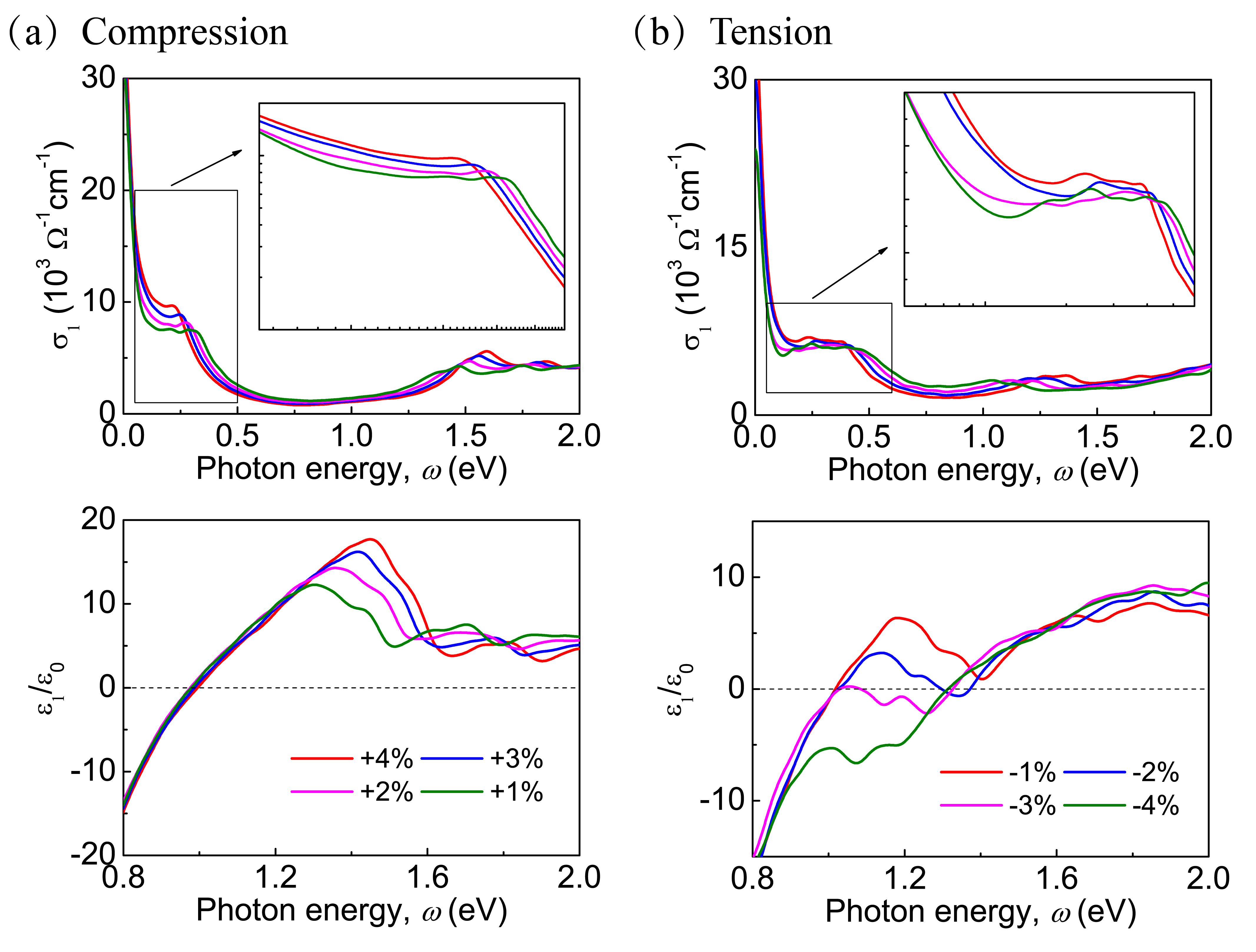}
	\caption{Real part of the optical conductivity $\sigma_{1}$ (upper panels) and dielectric function $\epsilon_{1}$ (lower panels) calculated for the in-plane direction under compressive (a) and tensile (b) strain as a function of the photon energy $\omega$. The inset figures use logarithmic scale. }
	\label{fig:strain_cond}
\end{figure}

Let us now discuss the strain-dependent electronic properties of ZrSiS. In Figure~\ref{fig:label5}(a), we show the band structures for the case of compressive and tensile uniaxial strain $u_{zz}$ of $1\%$, $3\%$, and $5\%$. 
One can see that the linear dispersion of states near the Fermi energy is unaffected by the uniaxial strain in the range from $-5\%$ to $+5\%$. The position of the Dirac points near the Fermi energy changes slightly, which is not expected to have any noticeable effects on the optical transitions at low energies. 
On the other hand, the position of the nonsymmorphic Dirac node at the X and R points is more susceptible to strain.   
As has been pointed out by Andreas \textit{et al.}, the location of these points in ZrSi$X$ ($X$=S, Se, Te) correlates strongly with the chemical pressure $c/a$ \cite{topp2016non}.
The most prominent effect of strain on the electronic structure of ZrSiS is the shift of the quadratic electron band along the energy axis. The tensile strain pushes this band toward the Fermi energy, while the compressive strain has the opposite effect. At around 2\% tensile strain, the electron states along the Z--R line cross the Fermi energy. 
The optical conductivity has contributions from both free carriers (Drude) and interband transitions in the vicinity of the Dirac points. The quadratic band, which crosses the nodal line at some k-points reduces the transition probability between the linear bands. This behavior is expected to have influence on the optical properties in the low-energy region.

Upon uniaxial compression of ZrSiS, its Fermi surface does not undergo any considerable modification, remaining topologically equivalent to the Fermi surface of pristine ZrSiS shown in Figure~\ref{fig:label1}(c). In contrast, in case of tensile strain the Fermi surfaces changes its topology as a consequence of the emerged conduction states with quadratic dispersion. One can distinguish between two Lifshitz transition occurring in stretched ZrSiS. When tensile stress reaches $P_{1}\sim1.3$ GPa, the previously disconnected hole pockets merge with each other, forming a ring at $k_z=\pi/c$. The corresponding merging region is highlighted in Figure~\ref{fig:label5}(b).
Up to $4\%$ tension, the electron and hole pockets are connected along the Z--R direction. When tensile stress reaches $P_{2} \sim 3.4$ GPa, a gap is being formed between the electron and hole pockets, manifesting itself the second transition in the Fermi surface topology [highlighted in Figure~\ref{fig:label5}(b)].
We also examine the nodal-line structure under tensile strain focusing at {\bf k}-points where Lifshitz transition takes place at the Fermi surface. The corresponding structure is shown in Figure~\ref{fig:label5}(c). Below $-2\%$ strain, the nodal lines form a continuous cage-like structure in the Brillouin zone. When tensile strain is increased, the nodal lines oriented in the $k_z$ direction get disconnected from the nodal loop at $k_{z}=\pm \frac{\pi}{c}$ along the Z--R direction at the cage corners. The corresponding separation between the nodal lines is increasing with strain, and can also be directly attributed to the appearance of the quadratic band along Z--R. Besides, one can see that the curvature of the nodal loop at $k_{z}=\pm \frac{\pi}{c}$ changes its sign when strain increases from $-1\%$ to $-4\%$.


We now examine the effect of strain on the interband screening.
To this end, we first calculate the unscreened plasma frequency shown in Figure~\ref{fig:strain_plasma} for the two crystallographic directions. While out-of-plane plasma frequency $\omega_{p,zz}$ exhibits a pronounced linear dependence as a function of strain, the in-plane plasma frequency, $\omega_{p,xx}$ demonstrates a more sophisticated dependence. Different behavior of $\omega_{p,zz}$ and $\omega_{p,xx}$ can be attributed to the difference in the Fermi velocities along the $x$- and $z$- directions. 
The strain-dependent screened plasma frequency $\omega^{scr}_{p}$ can be obtained from $\epsilon_{1}(\omega)$ shown in Figure~\ref{fig:strain_cond}. 
For compressive strain and small tensile strain up to $2\%$, $\omega^{scr}_{p}$ remains nearly a constant of around 1.0 eV. The situation for larger tensile strain is different. Due to the electronic Lifshitz transition, the nodal structure of $\epsilon_{1}(\omega)$ changes, leading to an enhancement of $\omega^{scr}_{p}$, which reaches $\sim$1.3 eV at $4\%$ tension.
The related interband screening $\epsilon_{\infty}=(\omega_p/\omega_p^{scr})^2$ is changing accordingly. As it is shown in Figure~\ref{fig:strain_plasma}, the in-plane component of $\epsilon_{\infty}$ is remaining around 9--10 up to 1\% tension, after which it decreases rapidly until the tension reaches 4\%, i.e. after the Fermi surface modification has occured. In this regime, $\epsilon_{\infty} \sim 3$--4, 
similar to the experimental values reported for ZrSiTe ($\sim 3.3$) \cite{ebad2019infrared}. This result is in favor of the chemical pressure mechanism proposed to describe the difference between the ZrSi$X$ ($X$=S, Se, Te) family members. Overall, the interband screening in moderately stretched ZrSiS is reduced considerably, which is expected to influence the optical response.

\begin{figure}[t]	
	\includegraphics[width=9.0cm]{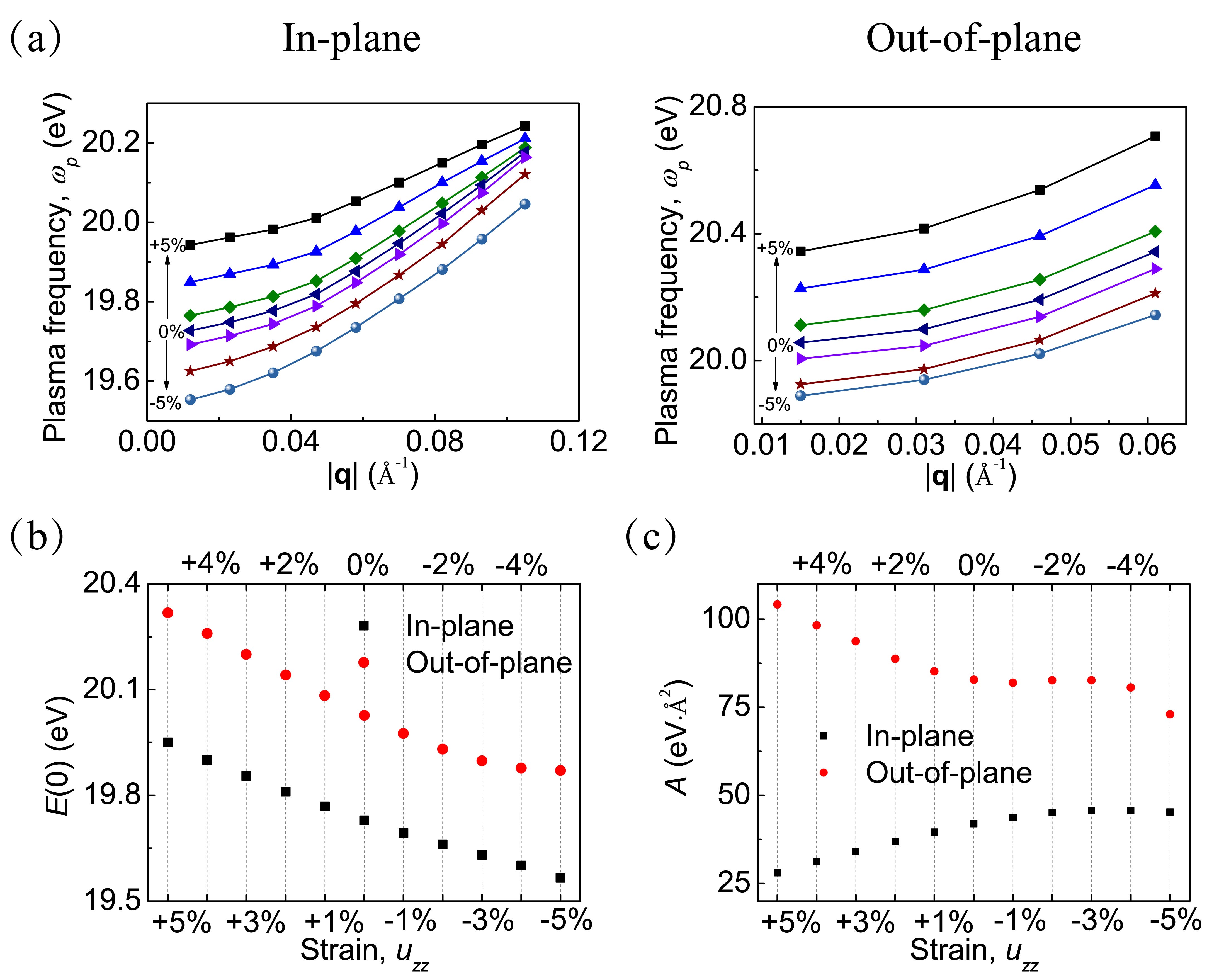}
	\caption{(a) Plasma frequency $\omega_p$ as a function of momentum ${\bf q}$ calculated for uniaxially strained ZrSiS along the in-plane (left) and out-of-plane (right) directions; (b) Plasma frequency in the long-wavelength limit (${\bf q}\to 0$) [fitted with Eq.~(\ref{dispersion})] as a function of strain; (c) Plasmon dispersion [fitted with Eq.~(\ref{dispersion})] as a function of strain. }
	\label{fig:label6}
\end{figure}

The in-plane conductivity calculated for different values and types of strain is shown in Figure~\ref{fig:strain_cond} as a function of the photon energy. 
The frequency-independent conductivity region tends to narrow (broaden) as the compressive (tensile) strain is applied.
Besides, the spectral weight in the low-energy region almost linearly enhances with load, gaining $\sim$50\% at $5\%$ compression.
On the contrary, the tensile strain reduces the spectral weight, yet not monotonously. At $\sim$3\% tension the optical conductivity is dropped, which apparently associated with the reduction of the interband contribution to the dielectric screening discussed earlier.
The observed lowering of the spectral weight in stretched ZrSiS is in line with the smaller flat optical conductivity observed in ZrSiSe with a larger $c/a$ lattice parameter \cite{ebad2019chemical}. 

At a larger energy scale, the effect of strain is less pronounced in the optical properties. In the range from 0.5 eV to 1.2 eV the optical conductivity is redshifted upon compression, while at larger frequencies it is blueshifted. The opposite situation is observed for the case of tensile strain. At low energies, the optical conductivity is mainly determined by the transitions between the linear bands in the electronic structure, as well as by the details of the Fermi surface. At energies above 1 eV the transitions between the parabolic bands become important, whose position on the energy axis is largely dependent on strain.
As a consequence, the characteristic U-shape of the optical conductivity around 1 eV almost disappears for more than $4\%$ tensile strain. 

Finally, we would like to comment on the effect of strain on the high-energy plasma excitations in ZrSiS. As this energy region is almost unrelated to the Fermi surface properties, the corresponding effect is less significant. In Figure~\ref{fig:label6}, we show the dispersion of the high-energy plasmon mode, as well as the corresponding parameters entering Eq.(\ref{dispersion}). Although the plasma frequency almost linearly changes with strain, the effect does not exceed a few percent for 5\% strain. 
In contrast, the dispersion of the plasma excitations can be tuned effectively by the compressive strain. While the dispersion along the out-of-plane direction decreases with strain gaining 30$\%$ at $+5\%$, the opposite effect is observed along the in-plane direction.

\section{Conclusions}
Based on first-principles calculations, we have systematically studied optical properties of nodal-line semimetal ZrSiS in the presence of uniaxial strain. 
We find that the characteristic frequency-independent optical conductivity is robust with respect to external uniaxial compression of up to 10 GPa. The compressive strain increases the spectral weight at low energies, but leads to a narrowing the flat conductivity region. The case of tensile strain is found to be more interesting. Upon tensile stress of 2 GPa, the Fermi surface undergoes a Lifshitz transition, resulting in a weakening of the interband dielectric screening. As a result, the spectral weight in the infrared region is reduced. The results obtained for stretched ZrSiS correlate with the properties of ZrSiSe and ZrSiTe, materials with larger lattice constants $c/a$. We, therefore, confirm the chemical pressure mechanism proposed in Ref.~\cite{ebad2019chemical} to describe variability in the electronic and optical properties of the ZrSi$X$ ($X$=S, Se, Te) family of compounds.  
On the other hand, the uniaxial tensile stress up to 2 GPa could be applied experimentally by flexure-based four-point mechanical wafer bending setup \cite{suthram2006piezoresistance}.

In the high-energy region, we found one lossy and one lossless plasmon modes at $\sim$5 and $\sim$20 eV, respectively. Although the frequencies of these modes remain almost unchanged in the presence of strain of up to 5\%, their dispersion can be effectively tuned. 
Being a layered material, ZrSiS exhibits strongly anisotropic dielectric response between the in-layer and stacking directions. This gives rise to the possibility of existence of hyperbolic plasmons in ZrSiS. Our calculations show that the hyperbolic regime indeed may exist within a frequency range of 0.6 eV around $\sim$5 and $\sim$20 eV.  
Overall, our findings provide insights into the mechanism behind the formation of optical properties in nodal-line semimetals ZrSi$X$, and pave the way for further optical studies, particularly in the ultraviolet spectral range.
\\
\newline

\newpage
\begin{acknowledgements}
SY acknowledges financial support from the National Key R$\&$D Program of China (Grant No. 2018FYA0305800) and National Science Foundation of China (Grant No. 11774269). A.N.R. acknowledges travel support from FLAG-ERA JTC2017 Project GRANSPORT. Numerical calculations presented in this paper have been performed on a supercomputing system in the Supercomputing Center of Wuhan University.
\end{acknowledgements}

\bibliographystyle{achemso}
\bibliography{references}

\end{document}